\documentclass[conference,a4paper]{IEEEtran}

\usepackage{url}
\usepackage{tocloft}
\usepackage{setspace}
\usepackage{amsmath}
\usepackage{multicol}
\usepackage{amssymb}
\usepackage[utf8]{inputenc}
\usepackage[T1]{fontenc}
\usepackage[font={small}]{caption}
\usepackage{cite}
\usepackage{enumerate}
\usepackage[square, comma, numbers, sort&compress]{natbib}
\usepackage{subfig}
\usepackage{bm}
\usepackage{multirow}
\usepackage{tabularx}
\usepackage{subfig}

\usepackage{url}
\usepackage{tocloft}
\usepackage{setspace}
\usepackage{amsmath}
\usepackage{multicol}
\usepackage{amssymb}
\usepackage[utf8]{inputenc}
\usepackage[T1]{fontenc}
\usepackage[font={small}]{caption}
\usepackage{cite}
\usepackage{enumerate}
\usepackage[square, comma, numbers, sort&compress]{natbib}
\usepackage{subfig}
\usepackage{bm}
\usepackage{multirow}

\newtheorem{proposition}{Proposition}
\newtheorem{remark}{Remark}

\newcommand{\vect}[1]{\boldsymbol{#1}}

\makeatother

\providecommand{\theoremname}{Theorem}

\makeatother

\providecommand{\lemmaname}{Lemma}

\makeatother

\providecommand{\propositionname}{Proposition}

\makeatother

\providecommand{\corname}{Corollary}

\makeatother

\providecommand{\remname}{Remark}

\ifCLASSINFOpdf
\else
\fi
\ifCLASSINFOpdf
   \usepackage[pdftex]{graphicx}
   \usepackage{epstopdf}
\else
\fi

\begin{document}
	
	\pagenumbering{gobble}
	
	\title{Autonomous Reconfigurable Intelligent Surfaces Through Wireless Energy Harvesting}


\author{\IEEEauthorblockN{Konstantinos Ntontin\IEEEauthorrefmark{1}, Alexandros-Apostolos A. Boulogeorgos\IEEEauthorrefmark{2}, Emil Bj\"{o}rnson\IEEEauthorrefmark{3}, Dimitrios Selimis\IEEEauthorrefmark{4},\\  Wallace Alves Martins\IEEEauthorrefmark{1}, Sergi Abadal\IEEEauthorrefmark{5}, Angeliki Alexiou\IEEEauthorrefmark{2}, Fotis Lazarakis\IEEEauthorrefmark{4}, Steven Kisseleff\IEEEauthorrefmark{1}, \\ and Symeon Chatzinotas\IEEEauthorrefmark{1} 			
	}\\\IEEEauthorblockA{\IEEEauthorrefmark{1}SnT, SIGCOM,  University of Luxembourg, Luxembourg,\\ e-mail: \{kostantinos.ntontin, steven.kisseleff, wallace.alvesmartins, Symeon.Chatzinotas\}@uni.lu\\\IEEEauthorrefmark{2}Department of Digital Systems, University of Piraeus, Greece, e-mail: al.boulogeorgos@ieee.org, alexiou@unipi.gr\\\IEEEauthorrefmark{3}Link\"{o}ping University and KTH Royal Institute of Technology, Sweden, email: emilbjo@kth.se\\\IEEEauthorrefmark{4}Institute of Informatics and Telecommunications, National Centre for Scientific Research ``Demokritos'', Greece,\\ e-mail: \{dselimis, flaz\}@iit.demokritos.gr\\\IEEEauthorrefmark{5}NaNoNetworking Center in Catalunya, Universitat Politècnica de Catalunya, Spain, email: abadal@ac.upc.edu
}}
	
	\maketitle
	
	\begin{abstract}
	    In this paper, we examine the potential for a reconfigurable intelligent surface (RIS) to be powered by energy harvested from information signals. This feature might be key to reap the benefits of RIS technology's lower power consumption compared to active relays.
	    We first identify the main RIS power-consuming components and then propose an energy harvesting and power consumption model. Furthermore, we formulate and solve the problem of the optimal RIS placement together with the amplitude and phase response adjustment of its elements in order to maximize the signal-to-noise ratio (SNR) while harvesting sufficient energy for its operation. Finally, numerical results validate the autonomous operation potential and reveal the range of power consumption values that enables it.
	\end{abstract}

\section{Introduction}
\label{Introduction}

Data-rate demands have been increasing in an exponential fashion for several decades. To prevent a possible capacity crunch, one candidate solution that has been put forward for the upcoming 5G generation of communication networks is the migration to frequency bands in the millimeter-wave (mmWave) range \cite{Andrews_what_will_5G_Be}. However, the higher blockage susceptibility in those bands make the coverage patchy. To overcome this issue, active relaying and the use of passive reflectors, such as dielectric mirrors, have been proposed. 
However, the main drawback of active relaying is need for a dedicated power supply for amplification, while the drawback of passive reflectors is their limited impact on the coverage due to the inability to dynamically control the reflection angle \cite{Khawaja_mmWave_Passive_Reflectors}. A promising solution that combines the benefits of both technologies without their disadvantages has been brought forward by the paradigm of reconfigurable intelligent surfaces (RISs) \cite{Marco_Di_Renzo_Survey_RISs, Bjornson2020a, Kisseleff_RIS}. 

RISs are artificial structures consisting of a dielectric substrate that embeds conductive elements, named reflective units (RUs), of sub-wavelength size and distance between adjacent elements. Typical RUs are comprised of either dipoles, patches, or string resonators, indicatively. By properly tuning their impedance through the use of semiconductor components, such as positive-intrinsic-negative (PIN) diodes,  field-effect transistors (FETs), and radio-frequency micro-electromechanical systems (RF-MEMS), their amplitude and phase response, with respect to an impinging electromagnetic wave, can be altered. Hence, besides reflecting an impinging beam towards an arbitrary direction or point, they can also act as absorbers of the impinging electromagnetic energy. In addition, an amount of power is needed for their reconfigurability, but not during a constant reflection configuration \cite{Bjornson2020a}. This is the reason why the RIS operation has been characterized as \emph{nearly passive}. If the reconfiguration is infrequent, the power consumption is arguably smaller than when operating an active relay.

With respect to RIS deployment in communication networks, there has been an intensive investigation in various domains \cite{Marco_Di_Renzo_Survey_RISs}. In addition, when compared to active relaying, which also allows beamforming in an arbitrary direction, several works  showcase that sufficiently large RISs can in fact outperform their relay counterparts \cite{Bjornson2020a,Comparison_RIS_Relaying_Boulogeorgos, Renzo2020, Bjornson_Relaying}. 

\textbf{\textit{Motivation and contribution}}: An important question that the RIS's nearly-passive feature raises is whether they can be power-autonomous by covering their needs through wireless energy harvesting. The vast majority of RIS related works dealing with wireless power transfer employ the RISs for assisting the transfer of power to end users and not for powering the RISs \cite{Pan_2020_SWIPT},\cite{Wu_2020_SWIPT}. To the best of our knowledge, only \cite{Wirelessly_Powered_RISs_TVT} considers wirelessly powered RISs. However, the authors do not present corresponding RIS power consumption and energy harvesting models nor introduce the RIS electronic modules that drive its power consumption. This is essential towards the identification of the advances needed in ultra-low power electronics that can realize the vision of autonomous RISs.
Motivated by this, the contribution of this work can be summarized as follows:
\begin{itemize}
	\item We present a comprehensive RIS power consumption model that captures its main power-consuming electronic components.
	
	\item We propose an energy harvesting model that is used for extracting the RIS harvested power and for formulating the optimization problem of interest. In particular, we focus on the optimal RIS placement as well as the amplitude and phase response adjustment of the RUs for maximizing the end-to-end signal-to-noise ratio (SNR), subject to the harvested power being sufficient for RIS autonomous operation.
	
	\item An analytical solution is provided for the amplitude and phase response optimal values of the RUs.
	
	\item Through the simulations, we provide a range of average power consumption of the RIS electronics that guarantees its autonomous operation.
	
\end{itemize}
	
The rest of this work is structured as follows: In Section II, the system model together with the RIS power-consuming modules are presented, whereas in Section III the considered RU reflection coefficient model and proposed energy harvesting and power consumption models are introduced. In Section IV, firstly the SNR is computed, subsequently the RIS harvested power model is introduced and, finally, the problem of interest is formulated and its solution is provided. Numerical results are provided in Section V, whereas Section VI concludes this paper.

\section{System Model and Power-Consuming Modules}

In this section, we present the system model under consideration and identify the RIS modules that consume power.

\subsection{System Model}

\begin{figure}
	\centering
	{\includegraphics[width=\columnwidth]{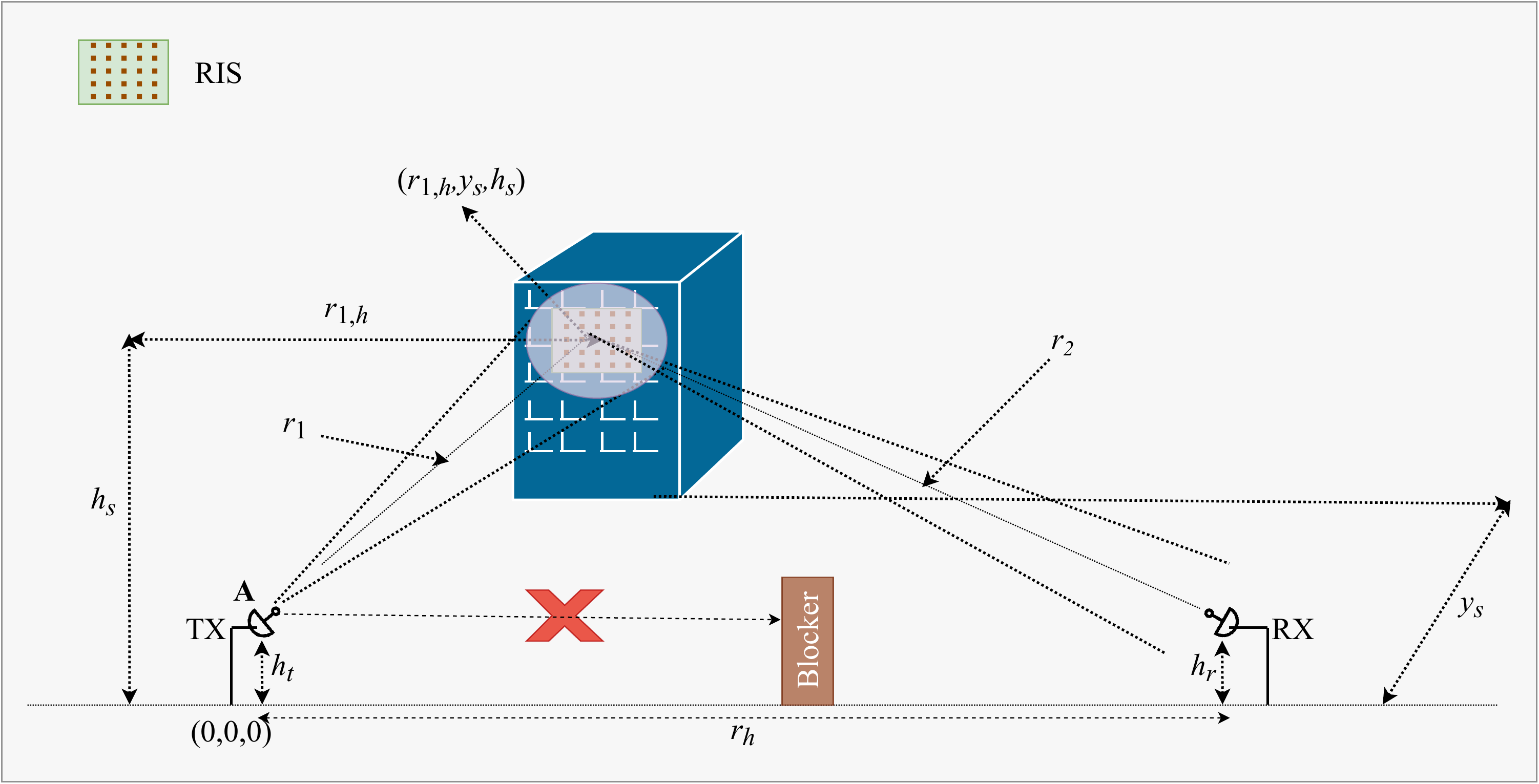}}
	\caption{Illustration of the system model and notation.}
	\label{Fig:Communication_through_an_RIS}	
\end{figure}

As illustrated in Fig.~\ref{Fig:Communication_through_an_RIS}, we consider a fixed-topology street-level scenario in which a transmitter (TX) communicates with a receiver (RX) through an RIS located in the far field of both the TX and RX. $r_{1,h}$, $r_{2,h}$, and $r_h$ denote the horizontal TX-RIS, RIS-RX, and TX-RX distances, respectively, while $h_t$, $h_{s}$, and $h_{r}$ are the TX, RIS, and RX heights, respectively. The incidence and departure angles of the electromagnetic wave with respect to the center of the illuminated area (RIS center) are respectively denoted by $\theta_i$ and $\theta_r$. The considered TX-RIS and RIS-RX blockage-free links are established in a mmWave band and constitute an alternative path to the direct TX-RX link that is assumed to be blocked. Such a scenario can be a typical future street-level fronthaul/backhaul mmWave setup. To countermeasure the high pathloss in such bands, both the TX and RX are equipped with highly directional antennas. Without loss of generality, we consider parabolic antennas with diameters $D_{t}$ and $D_{r}$, respectively. As a result, their maximum gain, denoted by $G_m^{max}$, $m\in\{t, r\}$, for $D_{t},D_{r} \gg \lambda$, where $\lambda$ represents the wavelength, is given by \cite{RIS_optimal_placement_Ntontin}
\begin{align}
	G_m^{max}=e_m\left(\frac{\pi D_m}{\lambda}\right)^2, \quad m\in\{t, r\},
\end{align}
where $e_m$ is their efficiency. Note that this type of antennas has been extensively used for wireless backhaul/fronthaul scenarios (see \cite{Backhaul_Fronthaul_Parabolic_Reflectors} and reference therein), due to their capability to support pencil-beamforming transmissions. 
In addition, we assume that the TX and RX antennas are pointing towards the center of illuminated RIS region. Furthermore, we note that even in such fixed-topology scenarios the RISs need to be occasionally reconfigured such as in the case of backhaul links in a mesh architecture \cite{Wireless_Mesh_Backhauling}.

We assume that the RIS has been deployed to have a line-of-sight path to the TX and RX. 
As far as the channel model is concerned, for both the TX-RIS and RIS-RX channels, we assume that the direct path dominates (as it has been reported through measurements in the mmWave bands \cite{Hanzo_MmWave_Channel_Model}) and use the free-space propagation model.


The RIS, consisting of $M_s=M_x\times M_y$ RUs of size $d_x \times d_y$, is configured to act as a beamformer, which, by proper adjustment of the RU phase response, is capable of steering an incident wave from any direction towards the angular direction $\theta_{\text{r}}$ to the RX direction. Each RU is an electrically-small low-gain element with gain pattern that can be expressed as \cite{RIS_optimal_placement_Ntontin}
\begin{align}
	\label{RIS_element_gain}
	G_{s}\left(\theta\right)= 4{\cos\left(\theta\right)}, \text{ with } 0\le\theta<\pi/2.
\end{align}Moreover, it is assumed that the transmission power is $P_t$ and that the received signal is subject to additive white complex Gaussian noise with power $\sigma^2$ \cite{RIS_optimal_placement_Ntontin}, computed in dBm as
\begin{align}
	\sigma^2=-174 + 10\log _{10} \left( W \right) + \mathcal{F}_{{\rm{dB}}},
\end{align}
where $\mathcal{F}_{{\rm{dB}}}$ is the noise figure in dB and $W$ is the bandwidth.
\begin{remark}
	The assumption of a fixed-topology scenario does not preclude the validity of the outcomes of this paper also for mobile scenarios. In particular, we can envisage a future urban scenario abundant in RISs that are mounted on fixed structures, such as buildings. In such a case, the equivalent question that can be answered by the results of this paper is which RIS should be chosen in order to maximize the SNR subject to the harvested power being sufficient for RIS autonomous operation. In such a case, we can still assume that the free-space TX-RIS-RX propagation model approximately applies considering the elevation of the RIS with respect to the positions of the TX and RX.  
\end{remark}

\subsection{RIS Power-Consuming Modules}

\label{Power_characteristics_semiconductor_components}

\subsubsection{Impedance-adjusting semiconductor components}

 This power consumption is characterized by two factors, namely the static power consumption and the dynamic power consumption. The first factor corresponds to their continuous power consumption due to leakage currents originating from the bias voltages when they operate in steady state. Usually,  the resulting direct-current (DC) power consumption is virtually negligible for FETs and RF MEMS. \cite{Paradigm_phase_shift}. On the other hand, the dynamic power consumption constitutes a non-negligible factor related to the charging and discharging of internal capacitors during bias voltage level changes needed for RU phase and amplitude response adjustment. It is present only when the semiconductor components change state, which means that its effect is alleviated in low-mobility scenarios.

\subsubsection{Energy-harvesting modules}

\label{Rectifying_circuits_characteristics}

For the RF-to-DC power conversion that is needed to power the RIS semiconductor components, we consider corporate feed networks in which the accumulated energy by a group of RUs is driven to a single rectifying circuit instead of dedicating one rectifying circuit per RU \cite{Amer_Metasurface_Energy_Harvesting}. The rectifying circuits can be either passive that exhibit negligible power consumption or can be active by incorporating active diodes that increase the conversion efficiency, but result in a non-negligible power consumption.

\subsubsection{Control network}

\label{Characteristics_of_the_control_network}

As described in \cite{Abadal_Programmable_Metamaterials}, the RIS needs to receive external commands regarding the configuration state it needs to assume. This can be achieved by either of two basic approaches: i) detached microcontroller architecture; ii) integrated architecture. In this paper, we consider the integrated architecture since it has a strong potential for enabling RIS autonomous operation due to expected low power consumption, as suggested by \cite{Abadal_Programmable_Metamaterials}. In such an architecture, the reconfiguration requests that are wirelessly received by the RIS are dispatched to an integrated network of communicating chips, which involve controllers that read the RU state and adjust the bias voltages of the impedance-adjusting semiconductor elements. The chip circuits that receive, interpret, and apply the commands exhibit their own static and dynamic power consumption due to leakage and transistor switching, respectively \cite{Tasolamprou_exploration_Intercell}. In addition, they are likely to use asynchronous logic due to the resulting small power consumption \cite{Asynchronous_Logic_Hypersurface}.

\section{RU reflection coefficient, energy-harvesting, and power consumption models}

In this section, we first present the considered model for the RU reflection coefficient and, subsequently, the energy-harvesting together with the power consumption model.

\subsection{RU Reflection-Coefficient Model}

We assume that the phase $\varphi_{p,l}$ and amplitude $A_{p,l}$ response of the $\left(p,l\right)$th RU can be tuned independently from each other. Although the RU phase and amplitude response are physically coupled \cite{Abeywickrama_TCOM_Practical_Phase_Shift}, there are design approaches that substantially alleviate such an inter-dependency  \cite{jia_broadband_2016}, \cite{Zhang2019}. Hence, the proposed model can serve as an upper bound of the expected performance.

\subsection{Energy Harvesting Model}

\label{Energy_harvesting_and_power_consumption_model}

\begin{figure}[!t]
	\label{Proposed_Energy_Harvesting_Model}
	\centering
	\includegraphics[width=.95\columnwidth]{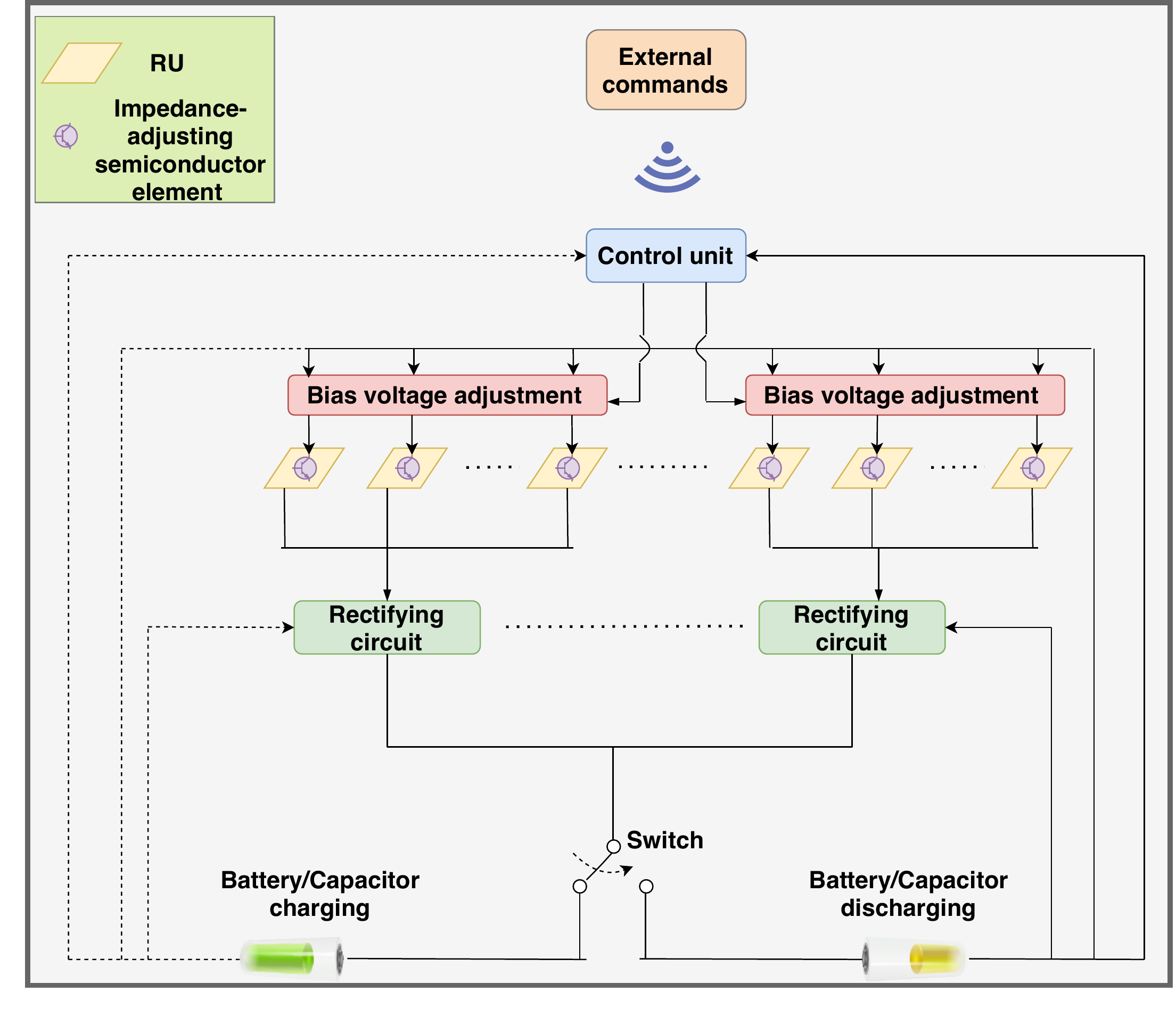}
	\caption{The proposed energy harvesting model.}
	\label{Proposed_Energy_Harvesting_Model}
\end{figure}

Our proposed energy harvesting model is depicted in Fig.~\ref{Proposed_Energy_Harvesting_Model}. We assume that each RU can act as both an energy harvester and as a reflector of the impinging electromagnetic radiation. The fraction of power that is absorbed (excluding the ohmic and other losses) is equal to $1-A_{p,l}^2$, while the fraction of reflected power is $A_{p,l}^2$. As depicted in Fig.~\ref{Proposed_Energy_Harvesting_Model}, the amount of RF energy that is absorbed by a group of RUs is converted into DC electricity through a rectifying circuit. The total DC energy output of the rectifying circuits charges one of two batteries/capacitors depicted in Fig.~\ref{Proposed_Energy_Harvesting_Model}. While one is charging, the other is discharging by providing power supply to both the controller chips that adjust the response of the RUs and the rectifying circuits. Once one of the batteries/capacitors is charged and the other one is discharged,\footnote{Assuming, ideally, the same charging and discharging rates.} the switch depicted in Fig.~\ref{Proposed_Energy_Harvesting_Model} swaps between them.

\subsection{Power Consumption Model}

For the RIS power consumption, denoted by $P_{\textrm{RIS}}$, by assuming one chip per RU it holds\footnote{The assumed linear dependency of the power consumption on the number of electronic chips ($M_sP_c$) can be considered as an upper bound on the amount that is expected in practice. This is due to the fact that the percentage of the electronic chips contributing to a steering angle change depends on the previous and targeted angle \cite{Workload_characterization_of_programmable_metasurfaces}.}
\begin{align}
P_{\textrm{RIS}}=M_sP_c+M_{\textrm{rect}}P_{\textrm{rect}},
\end{align} where with $P_c$ we incorporate both the power consumption of the control chip and the one of the impedance adjusting semiconductor component. Futhermore, $P_{\textrm{rect}}$ is the power consumption of each of the $M_{\textrm{rect}}$ rectifying circuits. We can consider $P_c$ as an equivalent continuous power consumption level (average value). For example, by denoting the static power consumption of the chip as $P_{\rm{static}}$, its dynamic one as $P_{\rm{dynamic}}$, and the percentage of time that the RIS needs to be reconfigured (this depends on the switching frequency and the reconfiguration duration) by $p_r$, it holds that $P_c=P_{\rm{static}}+p_rP_{\rm{dynamic}}$.

\section{Problem Formulation and Solution}

In this section, we first compute the SNR and, subsequently, we present the total harvested RIS power model. Finally, we formulate and solve the optimization problem of interest.

\subsection{Signal-to-Noise Ratio}

The SNR at the RX, which is given by \eqref{SNR_formula} at the top of the next page, can be obtained by following similar steps as in \cite[Appendix C]{RIS_optimal_placement_Ntontin}. The parameters $r_{1_{p,l}}$ and $r_{2_{p,l}}$ are the distances between the TX center and the $\left(p,l\right)$ element, and between the $\left(p,l\right)$ element, and the RX center, respectively, given by \eqref{distance_Tx_every_RIS_element}, shown at the top of the next page. $d_p$ and $d_l$ denote the distances between the RIS center and the $\left(p,l\right)$ RU in the x- and y-axis, respectively. Furthermore, $r_1$ and $r_2$ are the distances between the TX and the RIS center and between the RIS and RX center, respectively. They are given by
 
	\begin{figure*}
		\begin{small}
		\begin{align}
			\label{SNR_formula}
			\rho_s=\left(\frac{\lambda}{4\pi}\right)^4\frac{P_{t}G_{t}^{max}G_{r}^{max}G_{s}\left(\theta_i\right)G_{s}\left(\theta_r\right)}{r_1^2r_2^2\sigma^2}
			{\left| {\sum\limits_{p = 1}^{M_x} {\sum\limits_{l = 1}^{M_y} { {{	A_{p,l}}}\exp \!\left(\!\!-j\left( {{\!\varphi _{p,l}} \!+ \!\frac{{2\pi \left( {{r_{{1_{p,l}}}} \!+\! {r_{{2_{p,l}}}}} \right)}}{\lambda }} \right)\right)} } } \right|^2}.
		\end{align}\hrulefill\end{small}\end{figure*} 
\begin{figure*}
	\begin{small}
	\begin{equation}
		\label{distance_Tx_every_RIS_element}
		r_{1_{p,l}}\!=\!\sqrt{\left(\!r_{1,h}\!-\!d_p\right)^2\!\!+\!y_s^2+\!\!\left(h_s\!-\!h_{t}\!-\!d_l\right)^2},\;
		r_{2_{p,l}}\!=\!\sqrt{\left(\!r_{1,h}\!-\!d_p\right)^2+y_s^2\!+\!\left(h_s\!-\!h_{t}\!-\!d_l\right)^2}.
	\end{equation}
	\hrulefill
\end{small}
\end{figure*}
	\begin{small}
	\begin{equation}
		\label{distance_Tx_center_RIS}
		r_{1}\!=\!\sqrt{r_{1,h}^2\!\!+\!\!y_s^2\!\!+\!\!\left(h_s-h_{t}\right)^2},\; r_{2}\!=\!\sqrt{\left(r_h-r_{1,h}\right)^2\!+\!y_s^2\!+\!\left(h_s\!-\!h_{r}\right)^2}.
	\end{equation}
\end{small}
 Moreover, $\theta_i$ and $\theta_r$ are given by 
\begin{small}
\begin{align}
	\label{incidence_angle}
	\theta_i=\tan^{-1}\left(\frac{\sqrt{r_{{1},{h}}^2+\left(h_s-h_{t}\right)^2}}{y_s}\right),
\end{align}
\end{small}
\begin{small}
\begin{align}
	\label{departure_angle}
\theta_r=\tan^{-1}\left(\frac{\sqrt{\left(r_{{1},{h}}-r_h\right)^2+\left(h_s-h_{r}\right)^2}}{y_s}\right).
\end{align}
\end{small}

\subsection{RIS Harvested Power Model}

The absorbed power of the $\left(p,l\right)$ RU element, which we denote by $P_{{abs}_{p,l}}$, is given by
\begin{small}
\begin{align}
	\label{absorbed_power_per_RU}
	P_{\textrm{abs}_{p,l}}=\left(1-A_{p,l}^2\right)P_{{i}_{p,l}}=\left(\frac{\lambda}{4\pi}\right)^2
	\frac{P_{t}\left(1-A_{p,l}^2\right)G_{t}^{max}G_{s}\left(\theta_{i}\right)}{r_{1}^2}.
\end{align}
\end{small}
Hence, the total absorbed power of the RIS per communication time slot is the sum of $P_{\textrm{abs}_{p,l}}$ across all RUs. 
We let $\epsilon_{\textrm{conv}} \in \left(0,1\right) $ the RF-DC conversion efficiency, which is the same for all the employed rectifying circuits. The total harvested power from the RIS is then given by
\begin{align}
	P_{\textrm{harv}}=\epsilon_{\textrm{conv}}\sum\limits_{p =1}^{M_x} {\sum\limits_{l =1}^{M_y}}P_{\textrm{abs}_{p,l}}.
\end{align}
For enabling the perpetual (autonomous) operation of the RIS, it should hold that $P_{\textrm{harv}}\ge P_{\textrm{RIS}}$. By plugging
\eqref{RIS_element_gain},  \eqref{distance_Tx_center_RIS}, and \eqref{incidence_angle} into \eqref{absorbed_power_per_RU}, $P_{\textrm{harv}}$ as a function of $r_{1,h}$ and $\vect{A}=\left\{A_{p,l}\right\}$ is given by \eqref{Uncoupled_RU_amplitude_phase_response_harvested_power} at the top of the next page.
\begin{figure*}
	\begin{small}
	\begin{align}
		\label{Uncoupled_RU_amplitude_phase_response_harvested_power}
		P_{\textrm{harv}}\left(r_{1,h}, \vect{A}\right)=4\epsilon_{\textrm{conv}}\left(\frac{\lambda}{4\pi}\right)^2P_{t}G_{t}^{max}\sum\limits_{p = 1}^{M_x} {\sum\limits_{l = 1}^{M_y}}
		\frac{1-A_{p,l}^2}{r_{{1},{h}}^2+y_s^2+\left(h_s-h_{t}\right)^2}
		 \cos\left(\tan^{-1}\left(\frac{\sqrt{r_{{1},{h}}^2+\left(h_s-h_{t}\right)^2}}{y_s}\right)\right).
	\end{align}
	\hrulefill
\end{small}
\end{figure*}

\subsection{Problem Formulation and Solution}

The problem of interest is formulated as
\begin{align}
	\label{optimization_problem}
	&\underset{r_{1,h},\vect{A},\vect{\varphi}}{\text{maximize}}\;\rho_s\left(r_{1,h},\vect{A},\vect{\varphi}\right)\nonumber\\
	&\text{subject to}\;	P_{\textrm{harv}}\left(r_{1,h},\vect{A}\right)=P_{\textrm{RIS}}, \; 0<A_{p,l}<1,
\end{align}where $\vect{\varphi}=\left\{\varphi_{p,l}\right\}$ and $\rho_s\left(r_{1,h},\vect{A},\vect{\varphi}\right)$ is given by \eqref{SNR_formula_extended_uncoupled_case}, shown at the top of the next page, by plugging  \eqref{RIS_element_gain}, \eqref{distance_Tx_center_RIS}, \eqref{incidence_angle}, and \eqref{departure_angle}  into \eqref{SNR_formula}. As a constraint for the harvested power, we consider that it should be equal to the required amount needed to power the RIS electronics and not larger in order to devote more power to the information transmission.
	\begin{figure*}
		\begin{small}
		\begin{align}
			\label{SNR_formula_extended_uncoupled_case}
			\rho_s\left(r_{1,h},\vect{A},\vect{\varphi}\right)&=16P_tG_t^{max}G_r^{max}\left(\frac{\lambda}{4\pi}\right)^4\frac{\cos\left(\tan^{-1}\left(\frac{\sqrt{r_{{1},{h}}^2+\left(h_s-h_{t}\right)^2}}{y_s}\right)\right)\cos\left(\tan^{-1}\left(\frac{\sqrt{\left(r_{{1},{h}}-r_h\right)^2+\left(h_s-h_{r}\right)^2}}{y_s}\right)\right)}{\left(r_{{1},{h}}^2+y_s^2+\left(h_s-h_{t}\right)^2\right)\left(\left(r_{h}-r_{1,{h}}\right)^2+y_s^2+\left(h_s-h_{r}\right)^2\right)\sigma^2}\nonumber\\
			&\times{\left| {\sum\limits_{p = 1}^{M_x} {\sum\limits_{l =1}^{M_y} {\!\! A_{p,l}\exp \!\left(-j\left( {{\!\varphi _{p,l}} + \frac{{2\pi \left( {{r_{{1_{p,l}}}\left(r_{1,h}\right)} + {r_{{2_{p,l}}}\left(r_{1,h}\right)}} \right)}}{\lambda }} \right)\right)} } } \right|^2}.
		\end{align}
		\hrulefill
	\end{small}
	\end{figure*}

The main result of this paper is the following solution.

\vspace{2mm}


\begin{proposition}
	\label{Problem_solution}
	For the optimal values of $\varphi_{p,l}$, $A_{p,l}$, and $r_{1,h}$, denoted by $\varphi_{p,l}^*$, $A_{p,l}^*$, and $r_{1,h}^*$, respectively, it holds that
	\begin{small}
	\begin{align}
		\label{optimal_phase_repsonse_RU}
		\varphi_{p,l}^*=-\frac{{2\pi \left( {{r_{{1_{p,l}}}} \!+\! {r_{{2_{p,l}}}}} \right)}}{\lambda},
		\end{align}
	\end{small}
	\begin{small}
	\begin{align}
		\label{amplitude_RUs_optimal_value}
		A_{p,l}^*\!\!=\!\!\!\!\sqrt{\!\!1\!-\!\!\frac{P_{\textrm{RIS}}\left(\left(r_{1,h}^{*}\right)^2+y_s^2+\left(h_s-h_{t}\right)^2\!\!\right)}{4M_s\epsilon_{\textrm{conv}}\!\!\left(\!\frac{\lambda}{4\pi}\!\right)^2\!\!P_tG_t^{max}\!\!\cos\!\!\left(\!\!\tan^{-1}\!\!\!\left(\!\!\frac{\sqrt{\left(r_{{1},{h}}^{*}\right)^2+\left(h_s-h_{t}\right)^2}}{y_s}\!\!\right)\!\!\right)}}
		\end{align}
	\end{small}and $r_{1,h}^*$ is the value of $r_{1,h}$ that maximizes $G\left(r_{1,h}\right)$, given by \eqref{SNR_formula_extended_uncoupled_case_2}.
	\begin{figure*}
		\begin{small}
		\begin{align}
		\label{SNR_formula_extended_uncoupled_case_2}
			G\left(r_{1,h}\right)&=\frac{\cos\left(\tan^{-1}\left(\frac{\sqrt{r_{{1},{h}}^2+\left(h_s-h_{t}\right)^2}}{y_s}\right)\right)\cos\left(\tan^{-1}\left(\frac{\sqrt{\left(r_{{1},{h}}-r_h\right)^2+\left(h_s-h_{r}\right)^2}}{y_s}\right)\right)}{\left(r_{{1},{h}}^2+y_s^2+\left(h_s-h_{t}\right)^2\right)\left(\left(r_{h}-r_{1,{h}}\right)^2+y_s^2+\left(h_s-h_{r}\right)^2\right)\sigma^2}\nonumber\\
			&\times{\left(1-\frac{P_{\textrm{RIS}}\left(r_{1,h}^2+y_s^2+\left(h_s-h_{t}\right)^2\right)}{4M_s\epsilon_{\textrm{conv}}\left(\frac{\lambda}{4\pi}\right)^2P_tG_t^{max}\cos\left(\tan^{-1}\left(\frac{\sqrt{r_{{1},{h}}^2+\left(h_s-h_{t}\right)^2}}{y_s}\right)\right)}
			\right)}.
		\end{align}
		\hrulefill
	\end{small}
	\end{figure*}
	\end{proposition}
\begin{IEEEproof}
	The proof is provided in the appendix.
	\end{IEEEproof}

\section{Numerical results}

\begin{figure}
	\label{optimal_SNR_distance_vs_power_consumption}
	\centering
	{\includegraphics[width=\columnwidth]{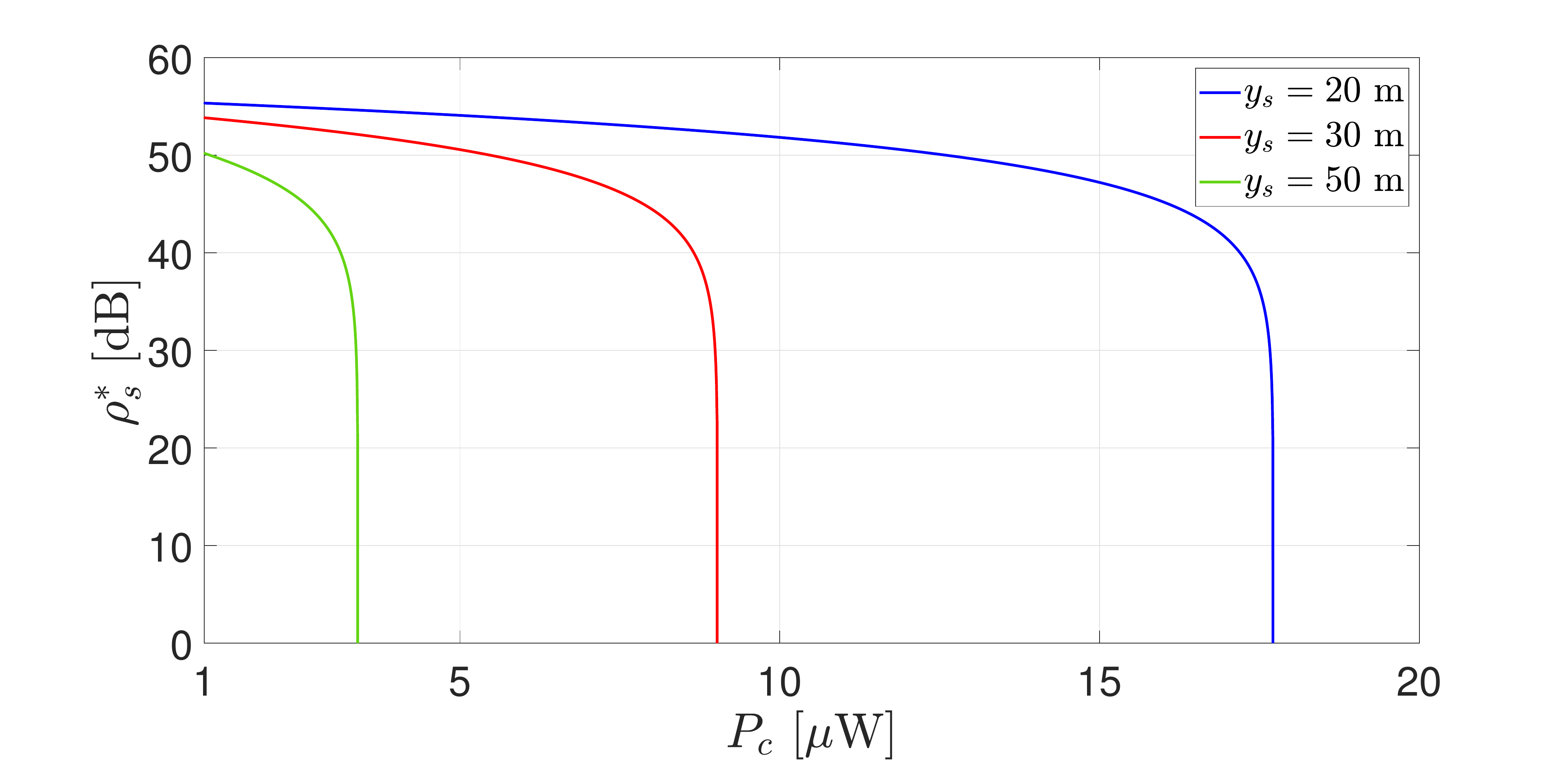}}\\
	(a) $\rho_s^{*}$ vs. $P_c$.
	{\includegraphics[width=\columnwidth]{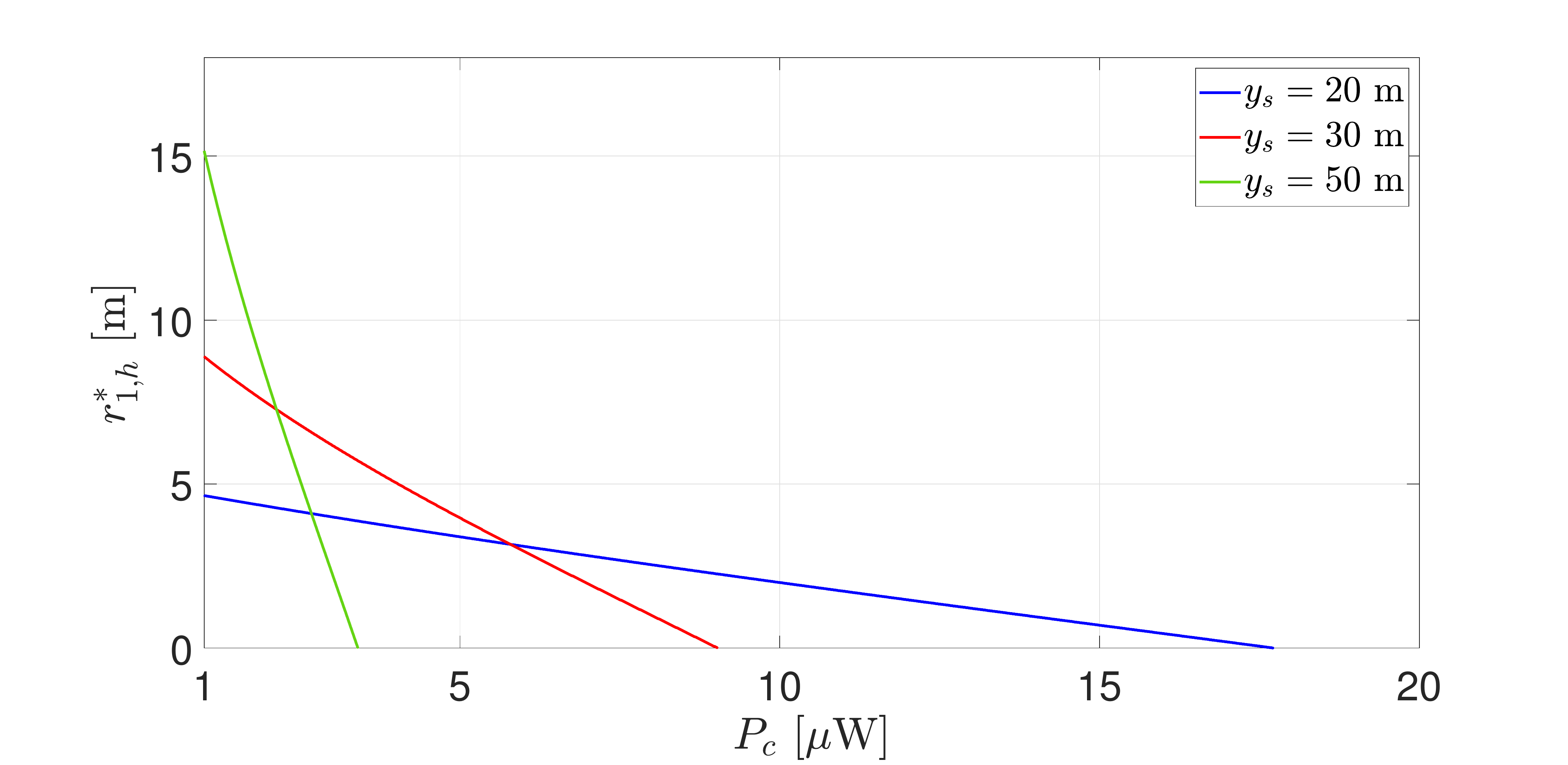}}\\
	(b)  $r_{1,h}^{*}$ vs. $P_c$.\\
	{\includegraphics[width=\columnwidth]{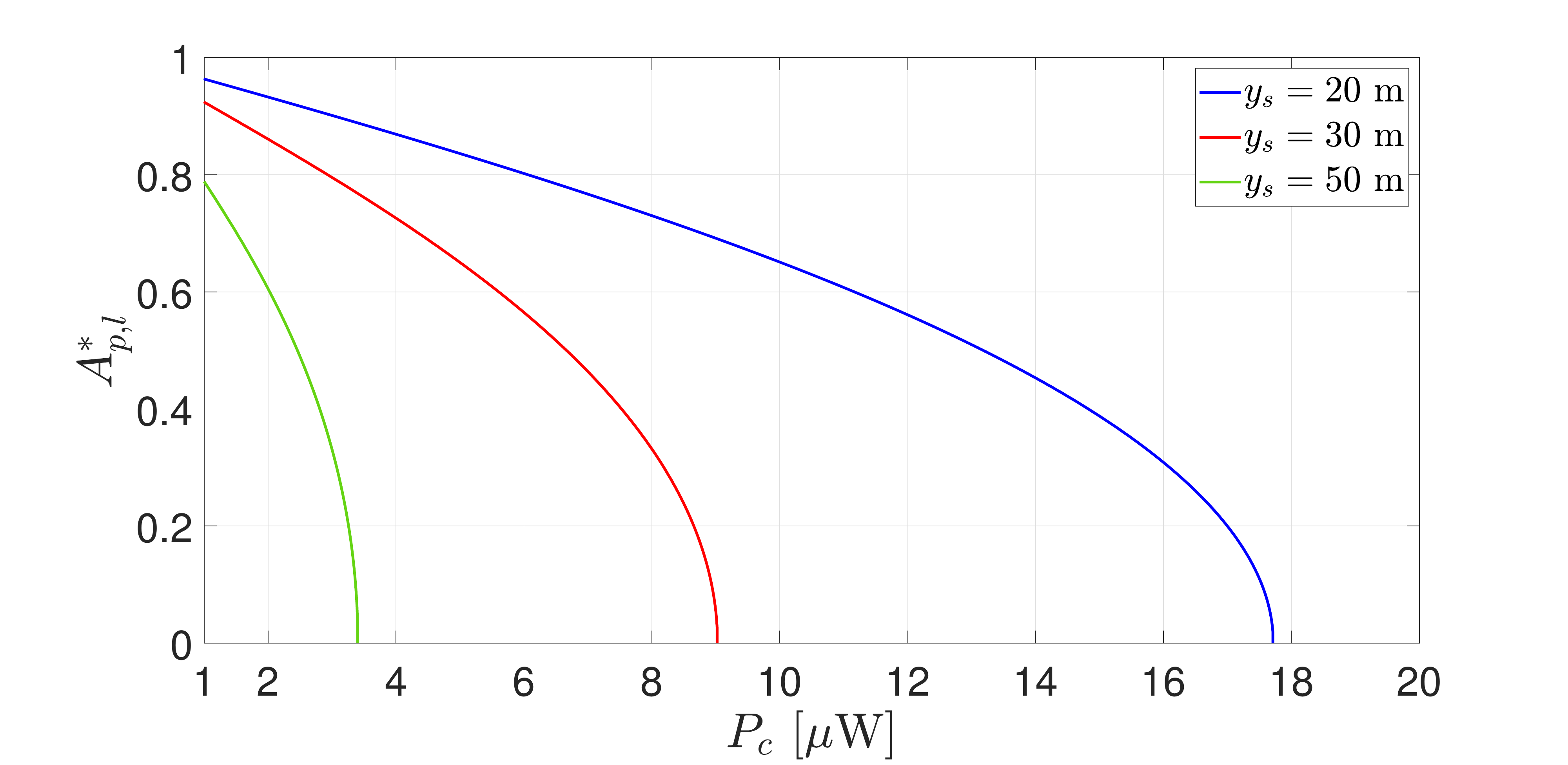}}\\
	(c)  $A_{p,l}^{*}$ vs. $P_c$.
	\caption{$\rho_s^{*}$, $r_{1,h}^{*}$, and $A_{p,l}^{*}$ vs. $P_c$.}
	\label{optimal_SNR_distance_vs_power_consumption}	
\end{figure}

We consider the parameter values of Table~\ref{Parameter_values}. In  Fig.~\ref{optimal_SNR_distance_vs_power_consumption}, we depict the optimal SNR, denoted by $\rho_s^{*}$, obtained by the optimal RIS placement, together with $r_{1,h}^{*}$ and $A_{p,l}^{*}$ versus $P_c$ for three values of $y_s$. From Fig.~\ref{optimal_SNR_distance_vs_power_consumption}(a), we observe that the higher $P_c$ is, the lower $\rho_s^{*}$ becomes until the RIS's power consumption cannot be covered by harvesting. In addition, we observe that as $y_s$ increases, the range of $P_c$ for which autonomous operation can be sustained is smaller. This is rational due to the higher TX-RIS distance. 

Furthermore, from Fig.~\ref{optimal_SNR_distance_vs_power_consumption}(b), we observe that the higher $P_c$ is, the closer to the TX the RIS needs to be placed. This trend is justified as follows: For relatively small values of $P_c$, the vast majority of the propagating energy should be dedicated to the SNR maximization since the RIS power needs can be covered by just a small amount of that energy, as it is verified by Fig.~\ref{optimal_SNR_distance_vs_power_consumption}(c). In such a case, the optimal RIS location could, depending on the configuration, be even closer to the middle of the TX-RX distance than the TX, especially for large $y_s$, as it is verified in \cite{RIS_optimal_placement_Ntontin}. On the other hand, for notable $P_c$ values, the RIS inevitably needs to be placed very close to the TX so that the highest possible amount of energy is harvested, as it is again verified by Fig.~\ref{optimal_SNR_distance_vs_power_consumption}(c).

\begin{table}[h]
	\label{Parameter_values}
	\caption{Parameter values used in the simulation.} 
	\centering 
	\scalebox{0.8}{
		\begin{tabular}{| c | c | c | c | } 
			\hline
			Parameter & Value & Parameter & Value\\[0.5ex]
			\hline
			\hline
			$f$& $28$ GHz& $d_x$, $d_y$& $\lambda/2$ \\[0.5ex] 
			\hline
			$P_{t}$& $1$ W & $M_{\textrm{rect}}$& $100$ \\ [0.5ex]
			\hline
			$P_{\textrm{rect}}$& 0 (passive rectification) & $h_{s}$& $12$ m \\ [0.5ex]
			\hline
			$r_{h}$& $100$ m& $h_{t}$, $h_r$& $3$ m \\ [0.5ex]
			\hline
			$W$& $2$ GHz & $D_{t}$, $D_r$& $30$ cm \\ [0.5ex]
			\hline
			$\mathcal{F}_{{\rm{dB}}}$& $10$ dB & $e_{t}$, $e_{r}$ & $0.7$\\[0.5ex]
			\hline
			$M_x$, $M_y$& $50$ & $e_{\textrm{conv}}$ & $0.6$ \\[0.5ex]
			\hline
	\end{tabular}}
	\label{Parameter_values} 
\end{table}

\section{Conclusions}

The case for RIS-aided communications, compared to active relaying, relies on the belief that the power consumption can be made lower. However, if the RIS still requires a wired power supply, the power reduction might be practically insignificant.
The purpose of this paper was to determine under which conditions, in terms of placement and element response, an autonomous RIS operation through energy harvesting from information signals is possible. The numerical results reveal that this is indeed possible if the average power consumption of the RIS electronic components does not exceed few microwatts. 
While the SNR over an RIS-aided communication link is the same when the TX and RX switch roles, the same does not apply for energy harvesting: the RIS should be close to the transmitter. 
The results from this study can help the system designer to identify the design requirements of future ultra-low power components in order to materialize such a vision.

\section*{Acknowledgements}

This work was supported by the European Commission's Horizon 2020 research and innovation programme ARIADNE (No. 871464) and the Luxembourg National Research Fund (FNR) under the CORE project RISOTTI.

\section*{Appendix}

We notice that  \eqref{SNR_formula_extended_uncoupled_case} is maximized when the complex terms in the norm are co-phased, which is achieved by \eqref{optimal_phase_repsonse_RU}. Hence, by plugging   \eqref{optimal_phase_repsonse_RU} into \eqref{SNR_formula_extended_uncoupled_case}, \eqref{optimization_problem} becomes

\begin{small}
	\begin{align}
		\label{optimization_problem_reformulation}
		&\underset{r_{1,h}, A_{p,l}}{\text{maximize}}\; F\left(r_{1,h}\right)\left(\sum\limits_{p=1}^{M_x} {\sum\limits_{l =1}^{M_y}}A_{p,l}\right)^2\nonumber\\
		&\text{subject to}\;	H\left(r_{1,h}\right)\sum\limits_{p = 1}^{M_x} {\sum\limits_{l = 1}^{M_y}}A_{p,l}^2=P_{\textrm{RIS}}, \; 0<A_{p,l}<1,
	\end{align}
\end{small}where $F\left(r_{1,h}\right)$ and $H\left(r_{1,h}\right)$ depend only on $r_{1,h}$ and can be extracted from \eqref{SNR_formula_extended_uncoupled_case} and \eqref{Uncoupled_RU_amplitude_phase_response_harvested_power}, respectively. We now define
\begin{small}
	\begin{align} 
		\Lambda\left(r_{1,h},\vect{A},\mu\right)&=F\left(r_{1,h}\right)\left(\sum\limits_{p = 1}^{M_x} {\sum\limits_{l = 1}^{M_y}}A_{p,l}\right)^2\nonumber\\
		&-\mu\left(H\left(r_{1,h}\right)\sum\limits_{p = 1}^{M_x} {\sum\limits_{l = 1}^{M_y}}A_{p,l}^2-P_{\textrm{RIS}}\right),
	\end{align}
\end{small}where $\mu$ is the Lagrange multiplier. By taking the first derivative of $\Lambda\left(r_{1,h},\vect{A}, \mu\right)$ with respect to each $A_{p,l}$ and equating it to zero, it holds that $A_{p,l}$ should be equal for each other, given by \eqref{amplitude_RUs_optimal_value} by replacing $r_{1,h}^*$ with $r_{1,h}$. Subsequently, by plugging $A_{p,l}$ into the objective function of \eqref{optimization_problem_reformulation}, \eqref{SNR_formula_extended_uncoupled_case_2} is obtained from which $r_{1,h}^*$ can be obtained by a linear search.



\bibliographystyle{IEEEtran}
\footnotesize{
	\bibliography{IEEEabrv,references}
}

\end{document}